
\newif\ifproofmode			
\proofmodefalse				

\newif\ifforwardreference		
\forwardreferencetrue			

\newif\ifeqchapternumbers		
\eqchapternumbersfalse			

\newif\ifsectionnumbers			
\sectionnumberstrue			

\newif\ifeqsectionnumbers		
\eqsectionnumbersfalse			

\newif\ifchaptersectionnumbers     	
\chaptersectionnumberstrue		

\newif\ifcontinuoussectionnumbers	
\continuoussectionnumbersfalse	

\newif\ifcontinuousnumbers		
\continuousnumbersfalse 		

\newif\iffigurechapternumbers		
\figurechapternumbersfalse		

\newif\ifcontinuousfigurenumbers	
\continuousfigurenumbersfalse		

\newif\ifcontinuousreferencenumbers     
\continuousreferencenumberstrue         

\newif\ifparenequations			
\parenequationstrue			

\newif\ifstillreading			

\font\eqsixrm=cmr6			
\def\marginstyle{\eqsixrm}		

\newtoks\chapletter			
\newcount\chapno			
\newcount\sectno			
\newcount\eqlabelno			
\newcount\figureno			
\newcount\referenceno			
\newcount\minutes			
\newcount\hours				

\newread\labelfile			
\newwrite\labelfileout			
\newwrite\allcrossfile			

\chapno=0
\sectno=0
\eqlabelno=0
\figureno=0


\def\chapternumberstrue{\eqchapternumberstrue}

%
\def\initialeqmacro{
    \ifproofmode
        \headline{\tenrm \today\ --\ \timeofday\hfill
                         \jobname\ --- draft\hfill\folio}
        \hoffset=-1cm
        \immediate\openout\allcrossfile=zallcrossreferfile
    \fi
    \ifforwardreference
        \openin\labelfile=zlabelfile
        \ifeof\labelfile
        \else
            \stillreadingtrue
            \loop
                \read\labelfile to \nextline
                \ifeof\labelfile
                    \stillreadingfalse
                \else
                    \nextline
                \fi
                \ifstillreading
            \repeat
        \fi
        \immediate\openout\labelfileout=zlabelfile
    \fi}


{\catcode`\^^I=9
\catcode`\ =9
\catcode`\^^M=9
\endlinechar=-1
\globaldefs=1


%
\def\chapfolio{			
    \ifnum \chapno>0 \relax
        \the\chapno
    \else
        \the\chapletter
    \fi}

%
\def\bumpchapno{
    \ifnum \chapno>-1 \relax
        \global \advance \chapno by 1
    \else
        \global \advance \chapno by -1 \setletter\chapno
    \fi
    \ifcontinuousnumbers
    \else
        \global\eqlabelno=0
    \fi
    \ifcontinuousfigurenumbers
    \else
        \global\figureno=0
    \fi
    \ifcontinuousreferencenumbers
    \else
        \global\referenceno=0
    \fi
    \sectno=0}

\def\bumpsectno{
    \global\advance\sectno by 1 \relax
    \ifeqsectionnumbers
        \ifcontinuoussectionnumbers
        \else
            \global\eqlabelno=0
        \fi
    \fi}

%
\def\setletter#1{\ifcase-#1 {}  \or\global\chapletter={A}
  \or\global\chapletter={B} \or\global\chapletter={C} \or\global\chapletter={D}
  \or\global\chapletter={E} \or\global\chapletter={F} \or\global\chapletter={G}
  \or\global\chapletter={H} \or\global\chapletter={I} \or\global\chapletter={J}
  \or\global\chapletter={K} \or\global\chapletter={L} \or\global\chapletter={M}
  \or\global\chapletter={N} \or\global\chapletter={O} \or\global\chapletter={P}
  \or\global\chapletter={Q} \or\global\chapletter={R} \or\global\chapletter={S}
  \or\global\chapletter={T} \or\global\chapletter={U} \or\global\chapletter={V}
  \or\global\chapletter={W} \or\global\chapletter={X} \or\global\chapletter={Y}
  \or\global\chapletter={Z}\fi}

%
\def\tempsetletter#1{\ifcase-#1 {}\or{} \or\chapletter={A} \or\chapletter={B}
 \or\chapletter={C} \or\chapletter={D} \or\chapletter={E}
  \or\chapletter={F} \or\chapletter={G} \or\chapletter={H}
   \or\chapletter={I} \or\chapletter={J} \or\chapletter={K}
    \or\chapletter={L} \or\chapletter={M} \or\chapletter={N}
     \or\chapletter={O} \or\chapletter={P} \or\chapletter={Q}
      \or\chapletter={R} \or\chapletter={S} \or\chapletter={T}
       \or\chapletter={U} \or\chapletter={V} \or\chapletter={W}
        \or\chapletter={X} \or\chapletter={Y} \or\chapletter={Z}\fi}

%
\def\chapshow#1{
    \ifnum #1>0 \relax
        #1
    \else
        {\tempsetletter{\number#1}\the\chapletter}
    \fi}

%
\def\today{\number\day\space \ifcase\month\or Jan\or Feb\or
        Mar\or Apr\or May\or Jun\or Jul\or Aug\or Sep\or
        Oct\or Nov\or Dec\fi, \space\number\year}

\def\timeofday{\minutes=\time    \hours=\time
        \divide \hours by 60
        \multiply \hours by 60
        \advance \minutes by -\hours
        \divide \hours by 60
        \ifnum\the\minutes>9 \relax
     		\the\hours:\the\minutes
 	\else
  		\the\hours:0\the\minutes
	\fi}


%
%
%
%
\def\chapnum{\bumpchapno \chapfolio}

\def\chaplabel#1{
    \ifforwardreference                             
        \write\labelfileout{                        
        \noexpand\expandafter\noexpand\def          
        \noexpand\csname CHAPLABEL#1\endcsname{\the\chapno}}
    \fi
    \global\expandafter\edef\csname CHAPLABEL#1\endcsname
    {\the\chapno}
    \ifproofmode
        \rlap{\hbox{\marginstyle #1\ }}
    \fi}

%
\def\sectnum{
    \bumpsectno
        \ifchaptersectionnumbers
            \chapfolio.
        \fi
    \the\sectno}

\def\sectlabel#1{
    \bumpsectno
    \ifforwardreference
        \immediate\write\labelfileout{
        \noexpand\expandafter\noexpand\def
        \noexpand\csname SECTLABEL#1\endcsname{\the\chapno.\the\sectno?!}}
    \fi
    \global\expandafter\edef\csname SECTLABEL#1\endcsname
    {\the\chapno.\the\sectno?!}	 			
    \ifproofmode
        \llap{\hbox{\marginstyle #1\ }}
    \fi
    \ifchaptersectionnumbers
        \chapfolio.
    \fi
    \the\sectno}

\def\sectref#1{                                  
    \ifundefined{SECTLABEL#1}                     
        ++                                        
        \ifproofmode
            \ifforwardreference
            \else
                \write16{ ***Undefined\space Section\space Reference #1*** }
            \fi
        \else
            \write16{ ***Undefined\space Section\space Reference #1*** }
        \fi
    \else
        \edef\LABxx{\getlabel{SECTLABEL#1}}
	\ifchaptersectionnumbers
            \def\LAByy{\expandafter\stripchap\LABxx}
	    \chapshow\LAByy.
	\fi
	\expandafter\stripsect\LABxx
    \fi
    \ifproofmode
        \write\allcrossfile{Section\space #1}
    \fi}

%
%
\def\eqnum{                                    
    \global\advance\eqlabelno by 1              
    \eqno(
    \ifeqchapternumbers
        \chapfolio.
    \fi
    \ifeqsectionnumbers
        \the\sectno.
    \fi
    \the\eqlabelno)}

\def\eqlabel#1{                                
    \global\advance\eqlabelno by 1              
    \ifforwardreference                     
        \immediate\write\labelfileout{\noexpand\expandafter\noexpand\def
        \noexpand\csname EQLABEL#1\endcsname
        {\the\chapno.\the\sectno?\the\eqlabelno!}}
    \fi
    \global\expandafter\edef\csname EQLABEL#1\endcsname
    {\the\chapno.\the\sectno?\the\eqlabelno!}
    \eqno(
    \ifeqchapternumbers
        \chapfolio.
    \fi
    \ifeqsectionnumbers
        \the\sectno.
    \fi
    \the\eqlabelno)
    \ifproofmode
        \rlap{\hbox{\marginstyle #1}}		
    \fi}

\def\eqalignnum{                               
    \global\advance\eqlabelno by 1              
    &(\ifeqchapternumbers
        \chapfolio.
    \fi
    \ifeqsectionnumbers
        \the\sectno.
    \fi
    \the\eqlabelno)}

\def\eqalignlabel#1{                   	
    \global\advance\eqlabelno by 1 	        
    \ifforwardreference                     
        \immediate\write\labelfileout{\noexpand\expandafter\noexpand\def
        \noexpand\csname EQLABEL#1\endcsname
        {\the\chapno.\the\sectno?\the\eqlabelno!}}
    \fi
    \global\expandafter\edef\csname EQLABEL#1\endcsname
    {\the\chapno.\the\sectno?\the\eqlabelno!}
    &(\ifeqchapternumbers
        \chapfolio.
    \fi
    \ifeqsectionnumbers
        \the\sectno.
    \fi
    \the\eqlabelno)
    \ifproofmode
        \rlap{\hbox{\marginstyle #1}}			
    \fi}

\def\dnum{                                     
    \global\advance\eqlabelno by 1              
    \llap{(	 				
    \ifeqchapternumbers
        \chapfolio.
    \fi
    \ifeqsectionnumbers
        \the\sectno.
    \fi
    \the\eqlabelno)}}

\def\dlabel#1{                                 
    \global\advance\eqlabelno by 1              
    \ifforwardreference                         
        \immediate\write\labelfileout{\noexpand\expandafter\noexpand\def
        \noexpand\csname EQLABEL#1\endcsname
        {\the\chapno.\the\sectno?\the\eqlabelno!}}
    \fi
    \global\expandafter\edef\csname EQLABEL#1\endcsname
    {\the\chapno.\the\sectno?\the\eqlabelno!}
    \llap{(
    \ifeqchapternumbers
        \chapfolio.
    \fi
    \ifeqsectionnumbers
        \the\sectno.
    \fi
    \the\eqlabelno)}
    \ifproofmode
        \rlap{\hbox{\marginstyle #1}}		
    \fi}

\def\eqref#1{\ifparenequations(\fi
    \ifundefined{EQLABEL#1}***
        \ifproofmode
            \ifforwardreference
            \else
                \write16{ ***Undefined\space Equation\space Reference #1*** }
            \fi
        \else
            \write16{ ***Undefined\space Equation\space Reference #1*** }
        \fi
    \else
        \edef\LABxx{\getlabel{EQLABEL#1}}
	\def\LAByy{\expandafter\stripsect\LABxx}
        \def\LABzz{\expandafter\stripchap\LABxx}
        \ifeqchapternumbers
            \chapshow{\LABzz}.
        \else
            \ifnum \number\LABzz=\chapno \relax
            \else
                \chapshow{\LABzz}.
            \fi
        \fi
	\ifeqsectionnumbers
	    \LAByy.
	\fi
        \expandafter\stripeq\LABxx
    \fi
    \ifparenequations)\fi
    \ifproofmode
        \write\allcrossfile{Equation\space #1}
    \fi}

%
\def\fignum{                                   
    \global\advance\figureno by 1\relax         
    \iffigurechapternumbers
        \chapfolio.
    \fi
    \the\figureno}

\def\figlabel#1{				
    \global\advance\figureno by 1\relax 	
    \ifforwardreference				
        \immediate\write\labelfileout{\noexpand\expandafter\noexpand\def
        \noexpand\csname FIGLABEL#1\endcsname
        {\the\chapno.\the\sectno?\the\figureno!}}
    \fi
    \global\expandafter\edef\csname FIGLABEL#1\endcsname
    {\the\chapno.\the\sectno?\the\figureno!}
    \iffigurechapternumbers
        \chapfolio.
    \fi
    \ifproofmode
        \llap{\hbox{\marginstyle #1\ }}\relax
    \fi
    \the\figureno}

\def\figref#1{					
    \ifundefined				
        {FIGLABEL#1}!!!!			
        \ifproofmode
            \ifforwardreference
            \else
                \write16{ ***Undefined\space Figure\space Reference #1*** }
            \fi
        \else
            \write16{ ***Undefined\space Figure\space Reference #1*** }
        \fi
    \else
        \edef\LABxx{\getlabel{FIGLABEL#1}}
        \def\LABzz{\expandafter\stripchap\LABxx}
        \iffigurechapternumbers
            \chapshow{\LABzz}.\expandafter\stripeq\LABxx
        \else \ifnum\number\LABzz=\chapno \relax
                \expandafter\stripeq\LABxx
            \else
                \chapshow{\LABzz}.\expandafter\stripeq\LABxx
            \fi
        \fi
        \ifproofmode
            \write\allcrossfile{Figure\space #1}
        \fi
    \fi}

%
%
\def\pagelabel#1{
    \ifforwardreference
        \write\labelfileout{
        \noexpand\expandafter\noexpand\def
        \noexpand\csname PGLABEL#1\noexpand\endcsname{\the\pageno}}
    \fi
    \global\expandafter\edef\csname PGLABEL#1\endcsname{\the\pageno}}

\def\pageref#1{
    \ifundefined
        {PGLABEL#1}***
        \ifproofmode
        \else
            \write16{ ***Undefined\space Page\space Reference #1*** }
        \fi
    \else
        \csname PGLABEL#1\endcsname
    \fi
    \ifproofmode
        \write\allcrossfile{Page\space #1}
    \fi}

%
\def\refnum{                                      
    \global\advance\referenceno by 1\relax         
    \the\referenceno}	                           

\def\internalreflabel#1{			
    \global\advance\referenceno by 1\relax 	
    \ifforwardreference				
        \immediate\write\labelfileout{\noexpand\expandafter\noexpand\def
        \noexpand\csname REFLABEL#1\endcsname
        {\the\chapno.\the\sectno?\the\referenceno!}}
    \fi
    \global\expandafter\edef\csname REFLABEL#1\endcsname
    {\the\chapno.\the\sectno?\the\figureno!}
    \ifproofmode
        \llap{\hbox{\marginstyle #1\hskip.5cm}}\relax
    \fi
    \the\referenceno}

\def\internalrefref#1{				
    \ifundefined				
        {REFLABEL#1}!!!!			
        \ifproofmode
            \ifforwardreference
            \else
                \write16{ ***Undefined\space Footnote\space Reference #1*** }
            \fi
        \else
            \write16{ ***Undefined\space Footnote\space Reference #1*** }
        \fi
    \else
        \edef\LABxx{\getlabel{REFLABEL#1}}
        \def\LABzz{\expandafter\stripchap\LABxx}
        \expandafter\stripeq\LABxx
        \ifproofmode
            \write\allcrossfile{Reference\space #1}
        \fi
    \fi}

%
\def\reflabel#1{\item{\internalreflabel{#1}.}}

%
\def\refref#1{\internalrefref{#1}}

\def\eq{\ifhmode Eq.~\else Equation~\fi}		
\def\eqs{\ifhmode Eqs.~\else Equations~\fi}

%
%
%
%

%
\def\getlabel#1{\csname#1\endcsname}
\def\ifundefined#1{\expandafter\ifx\csname#1\endcsname\relax}
\def\stripchap#1.#2?#3!{#1}			
\def\stripsect#1.#2?#3!{#2}			%
\def\stripeq#1.#2?#3!{#3}			
}  

\overfullrule = 0pt
\magnification = 1200
\baselineskip 21pt
\chapternumberstrue
\forwardreferencetrue
\initialeqmacro
\def\sh{\mathop{\rm sh}\nolimits}
\def\ch{\mathop{\rm ch}\nolimits}
\def\cth{\mathop{\rm cth}\nolimits}


\line{\hfill UMTG-169}

\vskip 0.2 in

\centerline{\bf Low-Temperature Thermodynamics}
\centerline{\bf of $A^{(2)}_2$ and $su(3)$-invariant Spin Chains}
\bigskip

\medskip

\centerline{Luca Mezincescu${}^*$,
Rafael I. Nepomechie\footnote*{Department of Physics, University of Miami,
Coral Gables, FL 33124, USA},
P. K. Townsend\footnote{$\dagger$}{DAMTP, University of Cambridge,
Silver Street, Cambridge, CB3 9EW, U.K.},
and A. M. Tsvelik\footnote{$\ddagger$}{Department of Physics, University
of Oxford, 1 Keble Road, Oxford, OX1 3NP, U.K.}}

\vskip 0.2 in

\bigskip

\centerline{\bf Abstract}

\vskip 0.2 in

We formulate the thermodynamic Bethe Ansatz (TBA) equations for the
closed (periodic boundary conditions) $A^{(2)}_2$ quantum spin chain in an
external magnetic field, in the (noncritical) regime where the anisotropy
parameter $\eta$ is real. In the limit $\eta \rightarrow 0$, we recover the
TBA equations of the antiferromagnetic $su(3)$-invariant chain in the
fundamental representation. We solve these equations for low temperature and
small field, and calculate the specific heat and magnetic susceptibility.

\vfill\eject

\noindent
{\bf \chapnum . Introduction and Summary}
\vskip 0.2truein

Given a quantum integrable lattice model in one space dimension, one can find
the eigenvalues of the Hamiltonian in terms of solutions of the model's Bethe
Ansatz (BA) equations. (See, e.g., Ref. \refref{qism}.)
Unfortunately, having found the eigenvalues, one is still quite far from
determining the model's physical properties. The main reason for this is that
the BA equations are in general very difficult to solve, in particular for $N$
(the number of lattice sites) finite. Considerable simplification occurs
in the thermodynamic ($N \rightarrow \infty$) limit. Provided one can formulate
a suitable ``string hypothesis'' for the solutions of the BA equations, the
problem is then to determine the densities $\rho_n (\lambda)$ and
$\tilde\rho_n (\lambda)$ of quasi-particles and quasi-holes, respectively.
In principle, this can be accomplished once one solves the so-called
thermodynamic Bethe Ansatz (TBA) equations for the quantities
$\epsilon_n = T \ln \left( \tilde\rho_n / \rho_n \right)$. Since the TBA
equations are an infinite set of coupled nonlinear integral equations, in
practice one solves them perturbatively (e.g., near $T=0$).

This program${}^{\refref{yangyang1}}$ has been successfully applied to a
number of integrable lattice models. Foremost among these are integrable
quantum spin chains -- e.g., the spin $1/2$ Heisenberg chain and its many
generalizations. The large body of
work${}^{\refref{takahashi1} - \refref{xxz/spin/s}}$
on the thermodynamics of quantum spin chains has had significant consequences
for both quantum field theory and condensed matter physics. (For a recent
introduction, see Ref. \refref{kingston}.)

In this paper, we focus on the closed (i.e., periodic boundary conditions)
$A^{(2)}_2$ chain, with Hamiltonian
$${\cal H} = \sum_{k=1}^{N-1} {\cal H}_{k,k+1} + {\cal H}_{N,1} \,,
\qquad\qquad {\cal H}_{k,k+1} = {d\over du} \check R_{k,k+1}(u)
\Big\vert_{u=0} \,.
\eqlabel{hamiltonian}  $$
Here
$$\check R(u) = {\cal P} R(u) \,, \eqnum $$
where ${\cal P}$ is the permutation matrix, and $R(u)$ is the $R$-matrix
associated with the twisted affine algebra $A^{(2)}_2$ in the fundamental
representation, which depends on the so-called anisotropy parameter $\eta$.
The Hilbert space is $\otimes^N V$, where $V$ is three dimensional.

This model was first explicitly constructed by Izergin and
Korepin${}^{\refref{izergin/korepin}}$.
The spectrum of the transfer matrix and the BA equations were first determined
using the analytical BA method${}^{\refref{reshetikhin1}}$, and later using the
algebraic BA method${}^{\refref{tarasov}}$. The
corresponding vertex model is equivalent to an $O(n)$ model on a
square${}^{\refref{nienhuis1}}$
or hexagonal${}^{\refref{reshetikhin2}}$ lattice. Such $O(n)$ models are
relevant${}^{\refref{nienhuis2}}$ to the study of polymers.

We distinguish two regimes: $\eta$ purely imaginary and $\eta$ purely real.
For zero external magnetic field, these correspond to critical and noncritical
regimes, respectively. By abuse of language, we shall refer to the two regimes
as ``critical'' and ``noncritical'' even for nonzero field.

For the $A^{(2)}_2$ chain in the critical regime, a general string hypothesis
has not yet been formulated. A new 2-string solution was found in
Ref. \refref{batchelor1},
and several new candidate 4-string solutions were found in Ref. \refref{miami}.
Presumably, there are new longer strings as well. In the absence of a
suitable string hypothesis, the TBA equations of the critical
$A^{(2)}_2$ chain cannot be formulated. An alternative approach
of investigating this model, based on finite-size corrections, has been
recently pursued by two groups${}^{\refref{devega1}, \refref{batchelor2}}$.
However, there is some disagreement between their results.

The difficulty in formulating a string hypothesis may be related to the
fact that, in the critical regime, the Hamiltonian of the
$A^{(2)}_2$ chain is {\it not} Hermitian. One might try to restrict the
space of states to a subspace in which the Hamiltonian is Hermitian.
However, since this model does not have a quantum-algebra symmetry,
it is not clear how to implement this restriction.

The situation for the open $A^{(2)}_2$ chain, whose Hamiltonian is the
same as \eqref{hamiltonian} except without the final term ${\cal H}_{N,1}$,
may be better. As shown in Ref. \refref{ijmp}, this model has the quantum
algebra symmetry $U_q [su(2)]$, and is integrable. Indeed, the spectrum of the
transfer matrix and the BA equations have been found, using a generalization
of the analytical BA method, in Ref. \refref{analytical}. In the critical
regime, one may be able to exploit the model's $U_q [su(2)]$ symmetry to make
suitable projections on the space of states, in analogy with the
$A^{(1)}_1$ case${}^{\refref{pasquier/saleur}}$.

As a warm-up exercise before addressing these problems, we consider in this
paper the closed $A^{(2)}_2$ chain in the {\it noncritical}\footnote*{in the
sense defined above} regime. Here the Hamiltonian is Hermitian, and there is no
difficulty in formulating${}^{\refref{miami}}$ a string hypothesis. We probe
this system at finite temperature $T$ in an external magnetic field $H$. In
analogy with the $A^{(1)}_1$
case${}^{\refref{johnson/mccoy}, \refref{takahashi2}}$, we expect various
phases in the $H - \Delta$ plane, where $\Delta \equiv \ch \eta$. In
particular,
for $\Delta \ge 1$ there should be a range of $H \ge 0$ for which the model
exhibits massless behavior.

We focus on the point $\Delta \rightarrow 1$, $H \rightarrow 0$
in the massless phase, where the model becomes
$su(3)$-invariant${}^{\refref{sutherland},\refref{kulish/reshetikhin}}$.
We solve the TBA equations in a systematic low-$T$ and small-$H$
perturbative expansion, along the lines of Johnson and
McCoy${}^{\refref{johnson/mccoy}}$.
These calculations employ Wiener-Hopf techniques, which were first used
in a similar context by Yang and Yang${}^{\refref{yangyang2}}$. However,
here we deal with a {\it system} of integral equations which requires
factorization of a {\it matrix} kernel${}^{\refref{gohberg/krein}}$.
We then compute the free energy $F(T,H)$, and determine the specific heat
and magnetic susceptibility.
Using the well-known relation${}^{\refref{cardy}, \refref{affleck1}}$ between
low-temperature specific
heat  and central charge ($c$) for critical systems, we arrive at the
value $c=2$ for the $su(3)$-invariant chain. This value coincides with that
obtained${}^{\refref{finite-size}}$ from finite-size corrections, and is
expected from the equivalence of this model to the level-one $su(3)$ WZW model
in the continuum limit. We believe that our value for the magnetic
susceptibility is new.

While there are other ways${}^{\refref{takahashi2}, \refref{filyov}}$
of calculating the low-temperature specific heat within the general TBA
approach, the method pursued here has the merit of treating this calculation
in the same manner as the one for the magnetic susceptibility.

We see no difficulty in computing thermodynamic quantities within the
TBA approach at other points in the $H - \Delta$ plane. However, we are
primarily interested in the massless phase, as this is where the connection
with field theory is better understood.

The outline of our paper is as follows. In Section 2, we formulate the TBA
equations for the closed $A^{(2)}_2$ chain in the noncritical regime, with
$\eta$ real. In Section 3, we take the limit $\eta \rightarrow 0$, and
arrive at the TBA equations for the antiferromagnetic $su(3)$-invariant chain
in the fundamental representation. In Section 4, we solve these equations for
small values of $T$ and $H$, and in Section 5 we calculate the free energy,
specific heat, and magnetic susceptibility. We present further discussion
of our results in Section 6.

\vskip 0.4truein

\noindent
{\bf \chapnum .  TBA equations for the $A^{(2)}_2$ chain}

\vskip 0.2truein

The Hamiltonian for the closed $N$-site $A^{(2)}_2$ chain
is given implicitly in terms of the $A^{(2)}_2$ $R$ matrix in
\eq\eqref{hamiltonian}. (Explicit expressions are given in Refs.
\refref{izergin/korepin}, \refref{reshetikhin1}, and \refref{ijmp}.) In an
external magnetic field $H$, the corresponding energy eigenvalues
are${}^{\refref{reshetikhin1}}$
$$E = - \sum_{k=1}^M {\sh^2 \eta \over \sin \eta (\lambda_k - {i\over 2})
\sin \eta (\lambda_k + {i\over 2})} - H \left( N - M \right)
\,, \eqlabel{eigenvalues} $$
where the real part of the complex variables $\lambda_k$ have
values in the interval $\left[ - \pi/2\eta \,, \pi/2\eta \right]$ and
satisfy the Bethe Ansatz (BA) equations
$$\eqalignno{
\left[{\sin \eta (\lambda_k + {i\over 2}) \over
\sin \eta (\lambda_k - {i\over 2})} \right]^N
= -& \prod_{j=1}^M {\sin \eta (\lambda_k - \lambda_j + i)
\cos \eta (\lambda_k - \lambda_j - {i\over 2} ) \over
\sin \eta (\lambda_k - \lambda_j - i)
\cos \eta (\lambda_k - \lambda_j + {i\over 2} )} \,, \cr
& \qquad\qquad\qquad\qquad k = 1, \cdots , M \,. \eqalignlabel{BA} \cr} $$
We consider the (noncritical) regime where $\eta$ is real and (without
loss of generality) positive.

We shall investigate the thermodynamics of this model, so we shall
need to solve the BA equations in the $N \rightarrow \infty$ and
$M \rightarrow \infty$ limit, with $M/N$ fixed. As is customary, we
adopt the ``string hypothesis'' which states that {\it all} the
solutions $\{ \lambda_k \,, k = 1, \cdots , \infty \}$ are collections
of $M_n$ ``strings'' of ``length'' $n$ of the form
$$\lambda_\alpha^{(n,l)} = \lambda_\alpha^n + i \left( {n+1\over 2} - l \right)
\,, \eqlabel{string} $$
where $l = 1, \cdots , n$; $\alpha = 0,1, \cdots, M_n$;
$n = 1, \cdots, \infty$;
and the ``centers'' $\lambda_\alpha^n$ are real. A particular solution of the
BA equations corresponds to a set of non-negative integers $\{M_n \}$
and the $M_n$ real numbers $\lambda_\alpha^n$ for each $n$. Observe that
the total number of $\lambda$ variables, and BA equations, is
$M = \sum_{n=1}^\infty n M_n$. These string solutions
are${}^{\refref{miami}}$ the same as those of the noncritical $A^{(1)}_1$
chain${}^{\refref{gaudin}}$.

It will prove convenient to introduce the functions
$$\eqalignno{
p_n(\lambda) &= i \ln \left[ {\sin \eta (\lambda + {i n\over 2}) \over
\sin \eta (\lambda - {i n\over 2})} \right] \,, \cr
q_n(\lambda) &= i \ln \left[ {\cos \eta (\lambda + {i n\over 2}) \over
\cos \eta (\lambda - {i n\over 2})} \right] \,, \eqalignnum \cr} $$
and the matrices
$$\Xi_{n m}(\lambda) = p_{n + m}(\lambda)
+ 2\sum_{l=1}^{min(n,m)-1} p_{|n - m| + 2l}(\lambda) + p_{|n - m|}(\lambda)
+ \sum_{l=1}^{min(n,m)} q_{2l - n - m -1}(\lambda) \,. \eqnum $$
We now substitute \eqref{string} into the BA equations \eqref{BA},
and take the product of the resulting equations for $\lambda_\alpha^{(n,l)}$
over the $n$ values of $l$. Taking the logarithm, we then obtain the
following equations for the centers $\lambda_\alpha^n$:
$$h^n (\lambda_\alpha^n) = J_\alpha^n \,,
\qquad\qquad \alpha = 0,1, \cdots, M_n \,; \qquad n = 1, 2, \cdots \,, \eqnum
$$
where
$$h^n (\lambda) = {1\over 2\pi} \left\{ N p_n(\lambda)
-\sum_{m=1}^\infty \sum_{\beta=0}^{M_{m}}  \Xi_{n m}
(\lambda - \lambda_\beta^{m}) \right\} \,, \eqlabel{hn} $$
and $J_\alpha^n$ are integers or half-integers. We make the conventional
assumption that $h^n(\lambda)$ is a monotonic increasing function of $\lambda$.
Let $J$ denote the set of
allowed values of $J_\alpha^n$, and $\tilde J$ its complement. If
$\lambda$ is such that $h^n (\lambda) \in J$, it is said to correspond
to a particle (of rapidity $\lambda$). If $\lambda$ is such that
$h^n (\lambda) \in \tilde J$, it is said to correspond to a hole.
Let $\rho_n (\lambda)$ be the density of particles and $\tilde\rho_n (\lambda)$
be the density of holes. Then
$$\rho_n(\lambda) + \tilde\rho_n(\lambda) = \lim_{N\rightarrow \infty}
{1\over N} {d\over d\lambda} h^n (\lambda) \,. \eqnum $$
An expression for the right hand side of this equation can be found by the
substitution of $N^{-1} \sum_\beta$ by
 $\int d\lambda' \rho(\lambda')$ in
\eqref{hn}. This leads to the equation
$$\tilde \rho_n +  \sum_{m=1}^\infty \left( A_{nm} + B_{nm} \right)
* \rho_{m} = a_n \,, \eqlabel{constraint}$$
where
$$\eqalignno{
A_{nm}(\lambda) &= \delta_{nm}\delta(\lambda)
+ \left(1 - \delta_{nm} \right) a_{|n - m|}(\lambda) + a_{n + m}(\lambda) \cr
& \qquad\qquad +2\sum_{l=1}^{min(n,m)-1} a_{|n - m| + 2l}(\lambda) \,,
\eqalignlabel{Anm} \cr
B_{nm}(\lambda) &=
\sum_{l=1}^{min(n,m)} b_{2l -n - m -1}(\lambda) \,,
\eqalignlabel{Bnm} \cr} $$
and
$$\eqalignno{
a_n(\lambda) &= {1\over 2\pi} {d\over d\lambda} p_n (\lambda)
= {\eta \over \pi} {\sh (\eta n) \over \ch (\eta n) - \cos (2\eta \lambda)} \,,
\eqalignlabel{an} \cr
b_n(\lambda) &= {1\over 2\pi} {d\over d\lambda} q_n (\lambda)
= {\eta \over \pi} {\sh (\eta n) \over \ch (\eta n) + \cos (2\eta \lambda)} \,,
\eqalignlabel{bn} \cr} $$
and $*$ denotes a convolution,
$$ \left( f * g \right) (\lambda) = \int_{-\pi/2\eta}^{\pi/2\eta}
d\lambda'\ f(\lambda - \lambda') g(\lambda') \,. \eqnum $$

For future reference we note here that
$$B_{nm}(\lambda) = - \left( s * A_{nm} \right) (\lambda + {\pi\over 2\eta})
\,, \eqlabel{future} $$
where $s(\lambda)$ is defined by
$$s(\lambda) = {\eta \over \pi}\sum_{k=-\infty}^\infty e^{-2i\eta k \lambda}
{1\over 2 \ch (\eta k)} \,. \eqlabel{s} $$
We also note that $s(\lambda)$, which can be expressed in terms of the
Jacobian elliptic function $dn$, has the property that
$$s*a_2 = a_1 - s \,, \qquad\qquad s*(a_{n+1} + a_{n-1}) = a_n
\qquad n>1 \,. \eqlabel{sproperty} $$
These and other relations which we give below can be easily derived with
the help of Fourier transforms, for which we use the following conventions
$$
f(\lambda) = {\eta \over \pi}\sum_{k=-\infty}^\infty e^{-2i\eta k \lambda}
\hat f_k \,, \qquad\qquad
\hat f_k = \int_{-\pi/2\eta}^{\pi/2\eta} d\lambda\ e^{2i\eta k \lambda}
f(\lambda)  \,. \eqnum $$

The thermodynamic limit of the energy per site is
$${E\over N} = -2\pi {\sh \eta\over \eta} \sum_{n=1}^\infty
\int_{-\pi/2\eta}^{\pi/2\eta} d\lambda\ a_n(\lambda) \rho_n (\lambda)
- H \left[ 1 - \sum_{n=1}^\infty n
\int_{-\pi/2\eta}^{\pi/2\eta} d\lambda\ \rho_n (\lambda) \right] \,,
\eqnum $$
while the entropy per site is
$${S\over N} = \sum_{n=1}^\infty
\int_{-\pi/2\eta}^{\pi/2\eta} d\lambda\ \left[
\left( \rho_n + \tilde \rho_n \right) \ln \left( \rho_n + \tilde \rho_n \right)
- \rho_n \ln  \rho_n - \tilde\rho_n \ln  \tilde\rho_n \right] \,. \eqnum $$
The equilibrium value of $\rho_n$ at temperature $T$ is
determined${}^{\refref{yangyang1}}$ by extremizing the free energy per site
$F/N = (E -TS)/N$. Note that the variation of $\tilde \rho_n$ is determined
in terms of the set of variations $\{ \delta \rho_n \}$ by the constraint
\eqref{constraint} which implies that
$$-{\delta \tilde\rho_n (\lambda)\over \delta \rho_{m} (\lambda')}
= A_{nm}(\lambda - \lambda') + B_{nm}(\lambda - \lambda') \,. \eqnum $$
Using this one finds that $F/N$ is extremized when the functions
$$\epsilon_n (\lambda) = T \ln \left({\tilde\rho_n (\lambda)\over
\rho_n (\lambda)}\right) \eqlabel{epsilon} $$
satisfy the thermodynamic Bethe Ansatz (TBA) equations
$$T \ln \left( 1 + e^{\epsilon_{n}/T} \right)
= \sum_{m=1}^\infty \left( A_{nm} + B_{nm} \right) * T \ln \left( 1 +
e^{-\epsilon_{m}/T} \right)
-2\pi{\sh \eta\over \eta} a_n + nH \,. \eqlabel{TBA} $$

Using the TBA equations one finds that, in equilibrium,
$${F \over N} = -T \sum_{n=1}^\infty \int_{-\pi/2\eta}^{\pi/2\eta} d\lambda\
a_n (\lambda) \ln \left( 1 + e^{-\epsilon_{n}(\lambda)/T} \right) - H
\,. \eqlabel{free1} $$
As in other integrable models, the free energy can be re-expressed as a
functional of only {\it one} of the $\epsilon_n(\lambda)$'s. To see this,
we introduce the matrix function
$$A^{-1}_{nm}(\lambda) = \delta(\lambda) \delta_{nm}
- s(\lambda) \left( \delta_{n, m+1} + \delta_{n, m-1} \right) \,,
\eqlabel{inverse} $$
where $s(\lambda)$ was defined in \eqref{s}. As the notation suggests,
it has the property
$$\sum_{n'=1}^\infty \left( A^{-1}_{nn'} * A_{n' m} \right) (\lambda)
= \delta(\lambda) \delta_{nm}  \,, \eqnum $$
which follows from \eqref{sproperty}. It has the further properties
$$\sum_{m=1}^\infty \left( A^{-1}_{nm} * a_m \right) (\lambda)
=  s(\lambda) \delta_{n 1} \,, \qquad\qquad
\sum_{m=1}^\infty A^{-1}_{nm} * m = 0 \,. \eqlabel{Anmproperty} $$
Inserting $A^{-1}A$ into
\eqref{free1} and using \eqref{Anmproperty} and \eqref{TBA} with
$B_{nm}$ replaced by the right hand side of \eqref{future}, one finds that
$${F\over N} = -2\pi {\sh \eta\over \eta}
\int_{-\pi/2\eta}^{\pi/2\eta} d\lambda\ a_1 (\lambda) r(\lambda)
- T \int_{-\pi/2\eta}^{\pi/2\eta} d\lambda\ r(\lambda)
\ln \left( 1 + e^{\epsilon_{1}(\lambda)/T} \right) \,. \eqlabel{free2} $$
Here
$$r(\lambda) = {\eta \over \pi}\sum_{k=-\infty}^\infty e^{-2i\eta k \lambda}
{\hat s_k\over 1 + (-1)^{k+1} \hat s_k} \,, \eqlabel{r} $$
where $\hat s_k = 1/2\ch \eta k$ are the Fourier coefficients of $s(\lambda)$.

{}From \eqref{free2} we see that in order to compute thermodynamic quantities
of our model, we need only determine $\epsilon_1(\lambda)$. Unfortunately,
the TBA equations \eqref{TBA} are coupled non-linear equations for all the
$\epsilon_n(\lambda)$. However, at low temperature these equations linearize
and a determination of $\epsilon_1(\lambda)$ becomes possible.

The type of low-temperature expansion depends crucially on the values of
the anisotropy parameter $\eta$ and the magnetic field $H$, as shown
for the $A^{(1)}_1$ model by Johnson and McCoy${}^{\refref{johnson/mccoy}}$.
We shall not pursue here a similar exhaustive analysis of the $A^{(2)}_2$
model.
Instead, we shall concentrate on the limits
$$\eta \rightarrow 0 \,, \qquad\qquad H \rightarrow 0 \,. $$
In the limit $\eta \rightarrow 0$, the $A^{(2)}_2$ model reduces to the
$su(3)$-invariant quantum spin chain in the fundamental
representation${}^{\refref{sutherland}, \refref{kulish/reshetikhin}}$. In the
following section, we shall explain how the TBA equations of the
$su(3)$-invariant model emerge from those of the $A^{(2)}_2$
model. We remark that the $su(3)$-invariant model allows the introduction of
two external (magnetic) fields because $su(3)$ has rank two. By our limiting
process we find only one combination.

\vskip 0.4truein
\noindent
{\bf \chapnum . The $\eta \rightarrow 0$ limit}
\vskip 0.2truein

We recall the $A^{(2)}_2$ BA equations \eqref{BA}:
$$\eqalignno{
\left[{\sin \eta (\lambda_k + {i\over 2}) \over
\sin \eta (\lambda_k - {i\over 2})} \right]^N
= -& \prod_{j=1}^M {\sin \eta (\lambda_k - \lambda_j + i)
\cos \eta (\lambda_k - \lambda_j - {i\over 2} ) \over
\sin \eta (\lambda_k - \lambda_j - i)
\cos \eta (\lambda_k - \lambda_j + {i\over 2} )} \,, \cr
& \qquad\qquad\qquad\qquad k = 1, \cdots , M \,. \eqalignlabel{again} \cr} $$
Consider a solution $\{ \lambda_1, \cdots, \lambda_M \}$. For
$\eta \rightarrow 0$, some of the $\lambda_k$'s remain finite. We call
these solutions $\lambda^{(1)}_k$, $k=1, \cdots, M^{(1)}$. The remaining
solutions become infinite; of these, we restrict ourselves
(following Ref. \refref{reshetikhin1}) only to those solutions whose behavior
as $\eta \rightarrow 0$ is given by
$\pm {\pi\over 2\eta} + \lambda^{(2)}_k$,  $k=1, \cdots, M^{(2)}$,
with $\lambda^{(2)}_k$ {\it finite}. Rewriting \eqref{again}
in terms of the new variables $\lambda^{(1)}_k$ and $\lambda^{(2)}_k$,
we obtain two families of BA equations,
which become in the $\eta \rightarrow 0$ limit
$$\eqalignno{
\left( {\lambda^{(1)}_k + {i\over 2} \over
        \lambda^{(1)}_k - {i\over 2} } \right)^N
= -& \prod_{j=1}^{M^{(1)}} { \lambda^{(1)}_k - \lambda^{(1)}_j + i \over
                             \lambda^{(1)}_k - \lambda^{(1)}_j - i }\
\prod_{j'=1}^{M^{(2)}} { \lambda^{(1)}_k - \lambda^{(2)}_{j'} - {i\over 2}
\over
                         \lambda^{(1)}_k - \lambda^{(2)}_{j'} + {i\over 2} }
\,, \quad\quad k = 1, \cdots , M^{(1)} \,, \cr
1 = -& \prod_{j=1}^{M^{(1)}} { \lambda^{(2)}_k - \lambda^{(1)}_j - {i\over 2}
                         \over \lambda^{(2)}_k - \lambda^{(1)}_j + {i\over 2}}\
      \prod_{j'=1}^{M^{(2)}} { \lambda^{(2)}_k - \lambda^{(2)}_{j'} + i \over
                               \lambda^{(2)}_k - \lambda^{(2)}_{j'} - i  }
\,, \quad\quad k = 1, \cdots , M^{(2)} \,.
\eqalignlabel{BAsu(3)} \cr} $$
These are precisely the BA equations for the $su(3)$-invariant
chain${}^{\refref{kulish/reshetikhin}}$.

In the formulas of the previous section, we must now distinguish
two classes of string centers. The string hypothesis becomes
$$\lambda_\alpha^{(n,r,l)} = \lambda_\alpha^{(n,r)}
+ i \left( {n+1\over 2} - l \right) \,, \eqnum $$
where $l = 1, \cdots , n$; $\alpha = 0,1, \cdots, M_n^{(r)}$; $r = 1, 2$;
$n = 1, \cdots, \infty$; and the centers $\lambda_\alpha^{(n,r)}$ are real.
The total number of $\lambda$'s of type $r$ is given by
$$M^{(r)} = \sum_{n=1}^\infty n M_n^{(r)} \,, \qquad\qquad r =1 , 2\,.
\eqnum $$
We correspondingly have densities of particles $\rho_n^{(r)}(\lambda)$
and holes $\tilde\rho_n^{(r)}(\lambda)$. The integral equations satisfied
by these densities are
$$\tilde \rho_n^{(r)} + \sum_{m=1}^\infty \sum_{s=1}^2
A_{nm} * C_{rs} * \rho_{m}^{(s)} = a_n \delta_{r 1} \,,
\eqlabel{constraintsu(3)} $$
where $A_{nm}$ and $a_n$ are the corresponding $\eta \rightarrow 0$ limits
of \eqref{Anm} and \eqref{an},
$$\eqalignno{
A_{nm}(\lambda) &= {1\over 2\pi} \int_{-\infty}^\infty d\omega\
e^{-i\lambda \omega} \left( \cth {|\omega|\over 2} \right) \left[
e^{-{|\omega|\over 2} |n - m|} - e^{-{|\omega|\over 2} (n + m)} \right]
\,, \eqalignnum \cr
a_n(\lambda) &= {1\over 2\pi} \int_{-\infty}^\infty  d\omega\
e^{-i\lambda \omega} e^{-n{|\omega|\over 2}}
= {1\over 2\pi} {n\over \lambda^2 + {n^2\over 4}}
\quad (n \ne 0) \,, \eqalignnum \cr} $$
$C_{rs}$ are the components of the $2 \times 2$ matrix
$$C(\lambda) = \left(
\matrix{\delta (\lambda) &-s(\lambda) \cr
             -s(\lambda) & \delta (\lambda) \cr} \right) \,, \eqlabel{Crs} $$
where
$$s(\lambda) = {1\over 2\pi} \int_{-\infty}^\infty  d\omega\
e^{-i\lambda \omega} {1\over 2 \ch {\omega\over 2}}
= {1\over 2 \ch \pi \lambda}
\,, \eqlabel{ssu(3)} $$
and $*$ now denotes the convolution
$$ \left( f * g \right) (\lambda) = \int_{-\infty}^\infty
d\lambda'\ f(\lambda - \lambda') g(\lambda') \,. \eqnum $$

The TBA equations are
$$T \ln \left( 1 + e^{\epsilon_n^{(r)}/T} \right)
= \sum_{m=1}^\infty \sum_{s=1}^2 A_{nm} * C_{rs}  * T \ln \left( 1 +
e^{-\epsilon_m^{(s)}/T} \right)
-2\pi a_n \delta_{r1} + n H \,, \eqlabel{TBAsu(3)} $$
where
$$\epsilon_n^{(r)} (\lambda) = T \ln \left({\tilde\rho_n^{(r)} (\lambda)\over
\rho_n^{(r)} (\lambda)}\right) \,.  \eqlabel{epsilonsu(3)} $$
Forming the convolution of the TBA equations with $A_{nm}^{-1}$ (which is given
by the expression \eqref{inverse}, with $s(\lambda)$ now given by
\eqref{ssu(3)}) and with $C_{rs}^{-1}$, we obtain
$$  T \ln \left( 1 + e^{-\epsilon_n^{(r)}/T} \right)
= \sum_{m=1}^\infty \sum_{s=1}^2 A_{nm}^{-1} * C_{rs}^{-1} * T \ln \left( 1 +
e^{\epsilon_m^{(s)}/T} \right)
+ 2\pi \delta_{n1} s^{(r)}  \,, \eqlabel{other} $$
where
$$s^{(1)}(\lambda) \pm s^{(2)}(\lambda) = {1\over 2\pi} \int_{-\infty}^\infty
d\omega\ {e^{-i\lambda \omega} \over 2 \ch {\omega\over 2} \mp 1}
= {2\over \sqrt 3} {\sh \left( (3 \pm 1) \pi \lambda / 3 \right)
\over \sh 2\pi \lambda} \,. \eqlabel{ss} $$
The equilibrium free energy is given by
$${F \over N} = -T \sum_{n=1}^\infty \int_{-\infty}^\infty d\lambda\
a_n (\lambda) \ln \left( 1 + e^{-\epsilon_n^{(1)}(\lambda)/T} \right) - H
\,. \eqlabel{free3} $$
With the help of \eqref{other}, this expression can be cast in a form
which depends only on $\epsilon_1^{(r)}(\lambda)$,
$${F\over N} = -2\pi \int_{-\infty}^\infty  d\lambda\  a_1 (\lambda)
s^{(1)}(\lambda)
- T \sum_{r=1}^2 \int_{-\infty}^\infty  d\lambda\ s^{(r)}(\lambda)
\ln \left( 1 + e^{\epsilon_1^{(r)}(\lambda)/T} \right) \,. \eqlabel{free4} $$
In the following sections, we shall solve the TBA equations \eqref{TBAsu(3)}
for small values of $T$ and $H$, and evaluate the expression
\eqref{free4} for the free energy.

\vfill\eject
\noindent
{\bf \chapnum . $T$ and $H$ expansion}
\vskip 0.2truein

We begin by rewriting the TBA equations of the
$su(3)$-invariant chain \eqref{TBAsu(3)} for $n=1$ in the form
$$\eqalignno{
T \ln \left( 1 + e^{\epsilon_1^{(r)}/T} \right)
=& \sum_{s=1}^2 A_{11} * C_{rs}
* T \ln \left( 1 + e^{-\epsilon_1^{(s)}/T} \right)
+ \sum_{m=2}^\infty \sum_{s=1}^2 A_{1m} * C_{rs}  * T \ln \left( 1 +
e^{-\epsilon_m^{(s)}/T} \right)  \cr
& \qquad -2\pi a_1 \delta_{r1} + H \,. \eqalignlabel{n=1} \cr}$$
We make the crucial assumption that $\epsilon_m^{(r)}(\lambda) > 0$
for $m>1$. (The results which we obtain below do not contradict this
assumption. For other examples, see Ref. \refref{tsvelick/wiegmann}.)
This means that $\exp (- \epsilon_m^{(r)}/T)$ goes to zero
exponentially as $T \rightarrow 0$ for $m>1$ and can therefore
be neglected.
On the other hand, $\epsilon_1^{(r)}(\lambda)$ can have either sign.
Defining $\varepsilon^{(r)}(\lambda)$ to be the $T \rightarrow 0$ limit of
$\epsilon_1^{(r)}(\lambda)$,
$$\varepsilon^{(r)}(\lambda) =
\lim_{T \rightarrow 0}\ \epsilon_1^{(r)}(\lambda) \,, \eqnum $$
we see that
$$ \lim_{T \rightarrow 0}\ T \ln \left( 1 + e^{\pm\epsilon_1^{(r)}/T} \right)
= \pm \varepsilon^{(r) \pm} \,, \eqlabel{limit} $$
where we use the standard notation
$$\varepsilon^- \equiv {1\over 2} \left( \varepsilon - |\varepsilon| \right)
\,, \qquad\qquad
\varepsilon^+ \equiv \varepsilon - \varepsilon^-  \,. \eqnum $$
It follows from \eqref{n=1} that
$$\varepsilon^{(r) +} = - \sum_{s=1}^2 A_{11} * C_{rs} * \varepsilon^{(s) -}
-2\pi a_1 \delta_{r1} + H \,. \eqlabel{almost} $$
It will be convenient for subsequent analysis to work with the following
equivalent equations, which do not involve $\varepsilon^{(r)-}$:
$$\eqalignno{
\varepsilon^{(1)} &= H -2\pi s^{(1)} + h * \varepsilon^{(1) +}
+ g * \varepsilon^{(2) +} \,, \cr
\varepsilon^{(2)} &= H -2\pi s^{(2)} + h * \varepsilon^{(2) +}
+ g * \varepsilon^{(1) +} \,, \eqalignlabel{transformed} \cr} $$
where $s^{(1)}$ and $s^{(2)}$ are defined in \eqref{ss}, and
$$ g = - s^{(1)} + s^{(2)} * a_1 \,, \qquad\qquad
   h = - s^{(2)} + s^{(1)} * a_1 \,. \eqlabel{h} $$
In order to obtain \eqref{transformed}, we substitute into \eqref{almost}
the explicit expressions for $A_{11}$ and $C_{rs}$ as given by Eqs.
\eqref{Anm} and \eqref{Crs}, as well as
$\varepsilon^{(r)-} = \varepsilon^{(r)} - \varepsilon^{(r)+}$; and then
we solve for $\varepsilon^{(r)}$ in terms of $\varepsilon^{(r)+}$ with the
help of Fourier transforms.

We observe for future reference that by a similar procedure,
\eqref{n=1} can be recast (after neglecting terms involving
$\epsilon_m^{(r)}$ with $m>1$, but before taking the $T \rightarrow 0$
limit of $\epsilon_1^{(r)}$) as follows:
$$\eqalignno{
\epsilon^{(1)}_1 &= H -2\pi s^{(1)}
+ h * T \ln \left( 1 + e^{\epsilon_1^{(1)}/T} \right)
+ g * T \ln \left( 1 + e^{\epsilon_1^{(2)}/T} \right) \,, \cr
\epsilon^{(2)}_1 &= H -2\pi s^{(2)}
+ h * T \ln \left( 1 + e^{\epsilon_1^{(2)}/T} \right)
+ g * T \ln \left( 1 + e^{\epsilon_1^{(1)}/T} \right) \,.
\eqalignlabel{transformed2} \cr} $$
Evidently, the $T \rightarrow 0$ limit of these equations gives
\eqref{transformed}.

We first briefly consider the case $H=0$, which corresponds to the ground
(vacuum) state. Since $s^{(1)}(\lambda)$ and
$s^{(2)}(\lambda)$ are positive for all $\lambda$, the equations
\eqref{transformed} have the solution
$$\eqalignno{
\varepsilon^{(1)}(\lambda) &= -2\pi s^{(1)}(\lambda) =
-{2\pi\over \sqrt 3} {\ch \left(  \pi \lambda/ 3 \right)
\over \ch (\pi \lambda)} \,, \cr
\varepsilon^{(2)}(\lambda) &= -2\pi s^{(2)}(\lambda) =
-{2\pi\over \sqrt 3} {\sh \left(  \pi \lambda/ 3 \right)
\over \sh (\pi \lambda)}
\,. \eqalignlabel{vacuum} \cr} $$
{}From the definition \eqref{epsilonsu(3)} and the constraint equations
\eqref{constraintsu(3)}, and also from the assumption that
$\epsilon_m^{(r)} > 0$ for $m>1$,
it follows that the ground-state densities of particles and holes are given by
$$\rho_1^{(1)}(\lambda) = {1\over \sqrt 3} {\ch \left(  \pi \lambda/ 3 \right)
\over \ch (\pi \lambda)} \,,  \qquad
\rho_1^{(2)}(\lambda) = {1\over \sqrt 3} {\sh \left(  \pi \lambda/ 3 \right)
\over \sh (\pi \lambda)} \,, \qquad
\rho_n^{(r)}(\lambda) = 0 \quad n >1  \,, $$
$$ \tilde \rho_n^{(r)}(\lambda) = 0  \,. \eqnum $$
That is, the ground state of the antiferromagnetic
$su(3)$-invariant chain consists of a ``condensate'' of strings of length 1.
Defining momenta $p^{(r)}(\lambda)$ of the quasi-particles by (see, e.g.,
Kirillov and Reshetikhin${}^{\refref{xxz/spin/s}}$)
$${d\over d\lambda}p^{(r)}(\lambda) = \varepsilon^{(r)}(\lambda) \,, \eqnum $$
we see from \eqref{vacuum} that for $\lambda \rightarrow \infty$, there is a
linear dispersion relation
$$\varepsilon^{(r)} = v_s p^{(r)} \,, \eqlabel{dispersion} $$
with the velocity of sound $v_s = 2\pi/3$.

We now turn to the case $H \ne 0$. For $H \rightarrow 0$, Eqs.
\eqref{transformed} can in principle be solved by iteration, with the
zeroth-order solution given by
$$\varepsilon^{(r)}_0(\lambda) = H -2\pi s^{(r)}(\lambda) \,. \eqnum $$
Since the functions $s^{(r)}(\lambda)$ are positive and monotonically
decreasing, for $H \rightarrow 0$ the functions $\varepsilon^{(r)}(\lambda)$
have a single zero. This zero is for $\lambda = O \left( \ln H \right)$.
Assuming an expansion in powers of $(\ln H)^{-1}$, we conclude that
$$ \varepsilon^{(r)}(\alpha^{(r)}) =  0
\eqnum $$
for
$$\alpha^{(r)} = -{3\over 2\pi} \left[
\ln \left( {{\sqrt 3}\over 2\pi} H \right) + \ln \kappa^{(r)} +
O \left( {1\over \ln H} \right) \right] \,, \eqlabel{kappa} $$
where the constants $\kappa^{(r)}$ (which are independent of $H$) have still
to be determined. We shall
further assume that the functions $\varepsilon^{(r)}(\lambda)$ have no
other zeros in the interval $\left( 0 \,, \infty \right)$. Hence,
$\varepsilon^{(r)}(\lambda) < 0$ for $\lambda$ in the interval
$\left( 0 \,, \alpha^{(r)} \right)$, and
$\varepsilon^{(r)}(\lambda) > 0$ for $\lambda$ in the interval
$\left( \alpha^{(r)} \,, \infty \right)$.

Observe that Eqs. \eqref{transformed} are temperature-independent. Using the
same approximations in the expression \eqref{free4} for the free energy this
too
will be temperature-independent. To find the leading order temperature
dependence we need to compute the leading correction to the
solutions $\epsilon^{(r)}_1 = \varepsilon^{(r)}$ of the linearized equations
\eqref{transformed}. In order to obtain this correction, we make the
substitution
$$\epsilon^{(r)}_1 (\lambda) = \varepsilon^{(r)}(\lambda)
+ \eta^{(r)} (\lambda) \eqlabel{expansion} $$
in \eqref{transformed2} and expand to leading order in $\eta^{(r)}$.
Since $\varepsilon^{(r)}$ is a solution of the linearized equations
\eqref{transformed}, we shall find inhomogeneous terms in the resulting
equations for $\eta^{(r)}$. Indeed, we have that
$$
\eta^{(1)} =
   h * \left\{ T \ln \left[ 1 +
e^{\left( \varepsilon^{(1)} + \eta^{(1)} \right)/T} \right]
- \varepsilon^{(1)+} \right\}
+  g * \left\{ T \ln \left[ 1 +
e^{\left( \varepsilon^{(2)} + \eta^{(2)} \right)/T} \right]
- \varepsilon^{(2)+} \right\} \,, \eqlabel{indeed}  $$
and the similar equation for $\eta^{(2)}$ is obtained by interchanging
the superscripts (1) and (2). Because $\varepsilon^{(r)}(\lambda)$ is
an even function of $\lambda$ with a single zero, at $\lambda = \alpha^{(r)}$,
for positive $\lambda$, we see that (assuming $\eta^{(r)}/T$ is small
and keeping terms linear in $\eta^{(r)}$)
$$\eqalignno{
& f * \left\{ T \ln \left[ 1 +
e^{\left( \varepsilon^{(r)} + \eta^{(r)} \right)/T} \right]
- \varepsilon^{(r)+} \right\} \cr
&= \left( \int_{-\infty}^{-\alpha^{(r)}} +
\int_{\alpha^{(r)}}^\infty \right) d\lambda'\ f(\lambda - \lambda')
\Big\{ T \ln \left[ 1 +
e^{\left( \varepsilon^{(r)}(\lambda') + \eta^{(r)}(\lambda') \right)/T} \right]
- \varepsilon^{(r)}(\lambda') \Big\} \cr
& \quad + \int_{-\alpha^{(r)}}^{\alpha^{(r)}} d\lambda'\ f(\lambda - \lambda')
T \ln \left[ 1 + e^{\left( \varepsilon^{(r)}(\lambda') + \eta^{(r)}(\lambda')
\right)/T} \right] \cr
& \approx \left( \int_{-\infty}^{-\alpha^{(r)}} +
\int_{\alpha^{(r)}}^\infty \right) d\lambda'\ f(\lambda - \lambda')
\eta^{(r)}(\lambda') + E_f^{(r)} \eqalignnum \cr}
$$
where the inhomogeneous term $E_f^{(r)}$ is given by
$$ E_f^{(r)} = f * T \ln \left( 1 + e^{-|\varepsilon^{(r)}|/T} \right)
\,. \eqnum $$
(Here $f$ represents either of the kernels $h$ and $g$ appearing in
\eqref{indeed}.)
For $T \rightarrow 0$, the major contribution to the integral comes from
the regions near the zeros of $\varepsilon^{(r)}$, so we expand
$\varepsilon^{(r)} (\lambda)$ about $\lambda = \alpha^{(r)}$,
$$ \varepsilon^{(r)} (\lambda)  = t^{(r)} \left( \lambda - \alpha^{(r)} \right)
+ O \left( (\lambda - \alpha^{(r)})^2 \right) \,, \qquad\qquad
t^{(r)} \equiv {d \varepsilon^{(r)} \over d\lambda}
\Big\vert_{\lambda = \alpha^{(r)}} \,. \eqlabel{tdefine} $$
One then finds that the leading $T$- dependence of $E_f^{(r)}$ is
$$\eqalignno{
E_f^{(r)}(\lambda) &= {2 T^2\over t^{(r)}} \left[ f(\lambda - \alpha^{(r)}) +
f(\lambda + \alpha^{(r)}) \right] \int_0^\infty du\ \ln \left( 1 + e^{-u}
\right) \cr
             &= {\pi^2 T^2 \over 6 t^{(r)}} \left[ f(\lambda - \alpha^{(r)}) +
f(\lambda + \alpha^{(r)}) \right] \,. \eqalignnum \cr} $$
Taking into account these $\eta$ - independent terms in the expansion of
\eqref{transformed2} to $O \left( \eta \right)$, we obtain the following
linear integral equation for $\eta^{(1)}$,
$$\eqalignno{
\eta^{(1)}(\lambda) =&  \left( \int_{-\infty}^{-\alpha^{(1)}} +
\int_{\alpha^{(1)}}^\infty \right) d\lambda'\ h(\lambda - \lambda')
\eta^{(1)}(\lambda') \cr
& + \left( \int_{-\infty}^{-\alpha^{(2)}} +
\int_{\alpha^{(2)}}^\infty \right) d\lambda'\ g(\lambda - \lambda')
\eta^{(2)}(\lambda') \cr
& + {\pi^2 T^2 \over 6 t^{(1)}} \left[ h(\lambda - \alpha^{(1)}) +
h(\lambda + \alpha^{(1)}) \right] \cr
& + {\pi^2 T^2 \over 6 t^{(2)}} \left[ g(\lambda - \alpha^{(2)}) +
g(\lambda + \alpha^{(2)}) \right] \,. \eqalignlabel{correction} \cr} $$
The similar expression for $\eta^{(2)}$ is obtained by interchanging the
superscripts $(1)$ and $(2)$ on $\eta$, $\alpha$ and $t$.
These equations for $\eta^{(r)}$ and those of \eqref{transformed} for
$\varepsilon^{(r)}$ complete our results for the $T$ - expansion of
$\epsilon^{(r)}_1$. We now turn to the expansion in powers of $\ln H$,
by which from Eqs. \eqref{transformed} and \eqref{correction} we shall
generate systems of integral equations of the Wiener-Hopf type.

It will prove convenient to work with the functions
$$S^{(r)}(\lambda) = \left\{ \matrix{e^{2\pi \alpha^{(r)}/3}
\kappa^{(r)} \varepsilon^{(r)} (\lambda + \alpha^{(r)})  & \lambda > 0 \cr
0 & \lambda < 0 \cr} \right. \,, \eqlabel{funcS} $$
and
$$T^{(r)}(\lambda) = \left\{ \matrix{ {6 e^{-2\pi \alpha^{(r)}/3}
\over \pi^2 T^2 \kappa^{(r)}}
\eta^{(r)} (\lambda + \alpha^{(r)})  & \lambda > 0 \cr
0 & \lambda < 0 \cr} \right. \,, \eqlabel{funcT} $$
instead of the functions $\varepsilon^{(r)}(\lambda)$ and
$\eta^{(r)}(\lambda)$.
The factors $e^{2\pi\alpha^{(r)}/3}$ and
$\left( e^{-2\pi \alpha^{(r)}/3}\right) /T^2$ in \eqref{funcS}
and \eqref{funcT}, respectively, are chosen such that the driving terms
in the equations for $S^{(r)}(\lambda)$ and $T^{(r)}(\lambda)$ have a
nonvanishing limit as $T \rightarrow 0$ and $H \rightarrow 0$. The
factors of $\kappa^{(r)}$, which were first introduced in \eqref{kappa}, appear
in \eqref{funcS} for a reason which will be explained below.

We return now to \eq\eqref{transformed}. We write the limits of
integration explicitly, keeping in mind that $\varepsilon^{(r)}(\lambda) > 0$
for $-\infty < \lambda <  -\alpha^{(r)}$ and for
$\alpha^{(r)} < \lambda < \infty $; and we shift the integration
variables so that they run from $0$ to $\infty$. Observe now that
$h\left( \lambda + 2\alpha^{(r)} \right)$ vanishes as $H \rightarrow 0$
($\alpha^{(r)} \rightarrow \infty$) for finite $\lambda$, and similarly
for $g\left( \lambda + \alpha^{(1)} + \alpha^{(2)} \right)$. The
functions $g\left( \lambda + \alpha^{(1)} - \alpha^{(2)} \right)$
and $g\left( \lambda + \alpha^{(2)} - \alpha^{(1)} \right)$ remain
finite, however, and survive in the $H \rightarrow 0$ limit of the
equations for $S^{(r)}(\lambda)$, which are (for $\lambda \ge 0$)
$$\eqalignno{
S^{(1)}(\lambda) &= {2\pi\over {\sqrt 3}}\left( 1 - \kappa^{(1)}
e^{-2\pi\lambda/3} \right) \cr
&+ \int_0^\infty d\lambda'\ \left[
h(\lambda - \lambda') S^{(1)}(\lambda') +
g(\lambda - \lambda' + \alpha^{(1)} - \alpha^{(2)}) S^{(2)}(\lambda')
\right] \,, \cr
S^{(2)}(\lambda) &= {2\pi\over {\sqrt 3}}\left( 1 - \kappa^{(2)}
e^{-2\pi\lambda/3} \right) \cr
&+ \int_0^\infty d\lambda'\ \left[
g(\lambda - \lambda' + \alpha^{(2)} - \alpha^{(1)}) S^{(1)}(\lambda') +
h(\lambda - \lambda') S^{(2)}(\lambda') \right] \,.
\eqalignnum \cr} $$

Similarly, from \eqref{correction}, we see that the $H \rightarrow 0$ limit
of the equations for $T^{(r)}(\lambda)$ are (for $\lambda \ge 0$)
$$\eqalignno{
T^{(1)}(\lambda) &= {h(\lambda)\over S^{(1)'}(0)} +
{g(\lambda + \alpha^{(1)} - \alpha^{(2)})\over S^{(2)'}(0)} \cr
&+ \int_0^\infty d\lambda'\ \left[
h(\lambda - \lambda') T^{(1)}(\lambda') +
g(\lambda - \lambda' + \alpha^{(1)} - \alpha^{(2)}) T^{(2)}(\lambda')
\right] \,, \cr
T^{(2)}(\lambda) &= {h(\lambda)\over S^{(2)'}(0)} +
{g(\lambda + \alpha^{(2)} - \alpha^{(1)})\over S^{(1)'}(0)} \cr
&+ \int_0^\infty d\lambda'\ \left[
g(\lambda - \lambda' + \alpha^{(2)} - \alpha^{(1)}) T^{(1)}(\lambda') +
h(\lambda - \lambda') T^{(2)}(\lambda') \right] \,,
\eqalignnum \cr} $$
where
$$ S^{(r) '}(0) \equiv
{d\over d\lambda}S^{(r)}(\lambda)\Big\vert_{\lambda = 0+}
= e^{2\pi \alpha^{(r)}/3} \kappa^{(r)} t^{(r)} \,. \eqnum $$

These equations can be written in the standard Wiener-Hopf form
$$\eqalignno{
S^{(r)}(\lambda) &= f_S^{(r)}(\lambda) + b_S^{(r)}(\lambda)
+ \sum_{s=1}^2 \int_{-\infty}^\infty d\lambda'\
K^{r s}(\lambda - \lambda')\ S^{(s)}(\lambda')  \,, \cr
T^{(r)}(\lambda) &= f_T^{(r)}(\lambda) + b_T^{(r)}(\lambda)
+ \sum_{s=1}^2 \int_{-\infty}^\infty d\lambda'\
K^{r s}(\lambda - \lambda')\ T^{(s)}(\lambda')   \,, \cr
&\qquad\qquad\qquad\qquad\qquad\qquad\qquad\qquad\qquad
-\infty < \lambda < \infty \,, \eqalignnum \cr} $$
where the kernels $K^{r s}(\lambda)$ are the components of the $2 \times 2$
matrix
$$K (\lambda) = \left(
\matrix{ h(\lambda) &  g(\lambda +\alpha^{(1)} -\alpha^{(2)}) \cr
         g(\lambda +\alpha^{(2)} -\alpha^{(1)}) &  h(\lambda) \cr}
\right) \,, \eqnum $$
and
$$f_S^{(r)}(\lambda) = \left\{ \matrix{
{2\pi\over {\sqrt 3}}\left( 1 - \kappa^{(r)} e^{-2\pi\lambda/3} \right)
& \lambda > 0 \cr
0 & \lambda < 0 \cr} \right. \,, \eqlabel{fS} $$
$$b_S^{(r)}(\lambda) = \left\{ \matrix{
0 & \lambda > 0 \cr
- \sum_{s=1}^2 \int_{-\infty}^\infty d\lambda'\
K^{r s}(\lambda - \lambda')\ S^{(s)}(\lambda')
& \lambda < 0 \cr} \right. \,, \eqlabel{bS} $$
and similarly
$$f_T^{(1)}(\lambda) = \left\{ \matrix{
{h(\lambda)\over S^{(1)'}(0)} +
{g(\lambda + \alpha^{(1)} - \alpha^{(2)})\over S^{(2)'}(0)}
& \lambda > 0 \cr
0 & \lambda < 0 \cr} \right. \,, \eqlabel{fT1} $$
$$f_T^{(2)}(\lambda) = \left\{ \matrix{
{h(\lambda)\over S^{(2)'}(0)} +
{g(\lambda + \alpha^{(2)} - \alpha^{(1)})\over S^{(1)'}(0)}
& \lambda > 0 \cr
0 & \lambda < 0 \cr} \right. \,, \eqlabel{fT2} $$
$$b_T^{(r)}(\lambda) = \left\{ \matrix{
0 & \lambda > 0 \cr
- \sum_{s=1}^2 \int_{-\infty}^\infty d\lambda'\
K^{r s}(\lambda - \lambda')\ T^{(s)}(\lambda')
& \lambda < 0 \cr} \right. \,. \eqlabel{bT} $$

We shall solve these equations by Fourier transform. We define the
Fourier coefficients of $S^{(r)}(\lambda)$ and $T^{(r)}(\lambda)$ by
$$\hat S^{(r)}(\omega) = \int_{-\infty}^\infty d\lambda\
e^{i \lambda \omega} S^{(r)}(\lambda) \,, \qquad\qquad
\hat T^{(r)}(\omega) = \int_{-\infty}^\infty d\lambda\
e^{i \lambda \omega} T^{(r)}(\lambda) \,. \eqnum $$
Since $S^{(r)}(\lambda)$ and $T^{(r)}(\lambda)$ vanish for $\lambda < 0$,
the functions $\hat S^{(r)}(\omega)$ and $\hat T^{(r)}(\omega)$ are
analytic in the upper-half-plane Im $\omega \ge 0$, which we denote by
$\Pi_+$. Observe that since
$S^{(r)}(0) = 0$ (as follows from $\varepsilon^{(r)}(\alpha^{(r)}) = 0$),
we have by contour integration
$$ S^{(r)}(0) =
-i \lim_{|\omega| \rightarrow \infty}\ \omega \hat S^{(r)}(\omega)
= 0  \,, \eqlabel{bc} $$
where the limit is taken in $\Pi_+$. We also note that
$$ S^{(r) '}(0) =
{d\over d\lambda}S^{(r)}(\lambda)\Big\vert_{\lambda = 0+}
= - \lim_{|\omega| \rightarrow \infty}\ \omega^2 \hat S^{(r)}(\omega)
\,. \eqlabel{second} $$
Care must be exercised in the derivation of this result because of the
discontinuity of the derivative of $S^{(r)}(\lambda)$ at $\lambda = 0$.

The Wiener-Hopf equations for $S^{(r)}$ in Fourier space are
$$\hat S^{(r)}(\omega) = \hat f_S^{(r)}(\omega) + \hat b_S^{(r)}(\omega)
+ \sum_{s=1}^2  \hat K^{r s}(\omega)\ \hat S^{(s)}(\omega)  \,,
\eqlabel{WHS} $$
where $\hat K^{r s}(\omega)$ are the components of the $2 \times 2$ matrix
$$\hat K (\omega) = \left(
\matrix{ \hat h(\omega) &  e^{-i \omega
\left( \alpha^{(1)}-\alpha^{(2)} \right)} \hat g(\omega)\cr
e^{i\omega \left( \alpha^{(1)}-\alpha^{(2)} \right)}
\hat g(\omega)  & \hat h(\omega) \cr} \right) \,.
\eqlabel{Kkernel} $$
Observe that (for $\omega$ and $\alpha^{(r)}$ real)
this matrix is Hermitian,
$$ \hat K (\omega)^\dagger = \hat K (\omega) \,. \eqnum $$
The factors of $\kappa^{(r)}$  in the definition \eqref{funcS} of
$S^{(r)}$ were chosen to arrange for this to be the case.

The Wiener-Hopf equations for $T^{(r)}$ in Fourier space are similarly
found to be
$$\hat T^{(r)}(\omega) = \hat f_T^{(r)}(\omega) + \hat b_T^{(r)}(\omega)
+ \sum_{s=1}^2  \hat K^{r s}(\omega)\ \hat T^{(s)}(\omega)  \,.
\eqlabel{WHT} $$

Since $\left( 1 - \hat K(\omega) \right)^{-1}$ is nonsingular, Hermitian and
positive-definite at $\omega = 0$, it is positive-definite for
$-\infty < \omega < \infty$. Thus, Theorem 8.2 of Gohberg and
Krein${}^{\refref{gohberg/krein}}$ implies that the following factorization
exists
$$ \left( 1 - \hat K(\omega) \right)^{-1} = G_+ (\omega)\ G_- (\omega)
\,, \qquad\qquad -\infty < \omega < \infty \,,
\eqlabel{factorization} $$
where $G_+ (\omega)$ and $G_+^{-1} (\omega)$ are analytic in $\Pi_+$ with
$G_+ (\omega) \rightarrow 1$ as $\omega \rightarrow \infty$ in $\Pi_+$,
and
$G_- (\omega)$ and $G_-^{-1} (\omega)$ are analytic in $\Pi_-$ with
$G_- (\omega) \rightarrow 1$ as $\omega \rightarrow \infty$ in $\Pi_-$.
Moreover, the fact $K (-\lambda) = K(\lambda)^T$ implies that
(for $\omega$ in $\Pi_-$)
$$G_-(\omega)^T = G_+(-\omega) \,, \eqlabel{Gproperty} $$
where the superscript $T$ denotes transpose. This lemma can be proved
from the formulas developed in Ref. \refref{gohberg/krein}.
As we shall see, explicit expressions for $G_+$ and $G_-$ are not needed
to compute the free energy to the order in which we work. A similar
phenomenon occurs in the Wiener-Hopf calculations of Yang and
Yang${}^{\refref{yangyang2}}$ and Johnson and
McCoy${}^{\refref{johnson/mccoy}}$.

Using the factorization \eqref{factorization}, the Wiener-Hopf equation
\eqref{WHS} for $\hat S^{(r)}$ can be rewritten (in $2 \times 2$ matrix
notation) as
$$G_+^{-1}\ \hat S = G_- \left( \hat f_S + \hat b_S \right) \,.
\eqlabel{interS} $$
Observe that the left hand side is analytic and bounded in $\Pi_+$,
whereas $G_- \hat b_S$ is analytic and bounded in $\Pi_-$. The term
$G_- \hat f_S$ has a decomposition as
$$G_- \hat f_S = P_- \left( G_- \hat f_S \right) +
P_+ \left( G_- \hat f_S \right) \,, \eqlabel{decomposition} $$
where $P_\pm \left( G_- \hat f_S \right)$ is analytic in $\Pi_\pm$.
This decomposition is uniquely specified by the requirement
$P_\pm \left( G_- \hat f_S \right) \rightarrow 0$ for $\omega \rightarrow
\infty$ in $\Pi_\pm$.
Taking the $P_+$ projection of \eqref{interS}, we have that
$$\hat S = G_+\ P_+ \left( G_- \hat f_S \right) \,. \eqlabel{formalS} $$
{}From \eqref{fS} we compute that
$$\hat f^{(r)}_S (\omega) = {2\pi i\over {\sqrt 3}}\left(
{1\over \omega + i\epsilon} - {\kappa^{(r)}\over \omega + {2\pi i/ 3}}
\right) \,, \eqnum $$
where one is to take $\epsilon \rightarrow 0$ at the end.
The decomposition \eqref{decomposition} of $G_- \hat f_S$ is then found
by subtracting the residues of $\hat f_S$, i.e.,
$$\eqalignno{
G_-(\omega) \hat f_S (\omega) =& {2\pi i\over {\sqrt 3}} \left\{
{1\over \omega + i\epsilon}
\left( G_-(\omega) - G_-(-i\epsilon) \right) \left( \matrix{ 1 \cr
                                                             1 \cr} \right)
-{1\over \omega + 2\pi i/3}
\left( G_-(\omega) - G_-(-2 \pi i/3) \right)
\left( \matrix{ \kappa^{(1)} \cr
                \kappa^{(2)} \cr} \right) \right\} \cr
&+ {2\pi i\over {\sqrt 3}} \left\{
{1\over \omega + i\epsilon} G_-(-i\epsilon)  \left( \matrix{ 1 \cr
                                                             1 \cr} \right)
-{1\over \omega + 2\pi i/3} G_-(-2 \pi i/3)
\left( \matrix{ \kappa^{(1)} \cr
                \kappa^{(2)} \cr} \right) \right\} \,. \eqalignnum \cr}$$
Hence,
$$\hat S(\omega) = {2\pi i\over {\sqrt 3}} {1\over \omega + i\epsilon}
G_+(\omega) G_-(0) \left( \matrix{ 1 \cr
                                   1 \cr} \right)
- {2\pi i\over {\sqrt 3}} {1\over \omega + 2\pi i/3}
G_+(\omega) G_-(-2 \pi i/3)
\left( \matrix{ \kappa^{(1)} \cr
                \kappa^{(2)} \cr} \right) \,. \eqnum $$
The boundary condition \eqref{bc} can now be seen to be equivalent to
the condition
$$ G_-(-2 \pi i/3)  \left( \matrix{ \kappa^{(1)} \cr
                                    \kappa^{(2)} \cr} \right)
= G_-(0)  \left( \matrix{ 1 \cr
                          1 \cr} \right)  \,,
\eqlabel{explicitkappa} $$
which can be solved for the parameters $\kappa^{(r)}$.
Using this result we conclude that $\hat S(\omega)$ is given by
$$\hat S(\omega) = {2\pi i\over {\sqrt 3}}
\left( {1\over \omega + i\epsilon} - {1\over \omega + 2\pi i/3} \right)
G_+(\omega) G_-(0) \left( \matrix{ 1 \cr
                                   1 \cr} \right)
 \,. \eqlabel{Ssolution} $$
{}From \eq\eqref{second} and the fact that
$G_+ (\omega) \rightarrow 1$ as $|\omega| \rightarrow \infty$ in $\Pi_+$,
we find that
$$ S^{(r) '}(0)
= {4\pi^2 \over 3 {\sqrt 3}}
G_-(0) \left( \matrix{ 1 \cr
                       1 \cr} \right)
\,. \eqlabel{derivative} $$

We turn now to the equation \eqref{WHT} for $\hat T(\omega)$.
Proceeding as before, we use the factorization \eqref{factorization}
to arrive at the formal solution
$$\hat T = G_+\ P_+ \left( G_- \hat f_T \right)  \eqlabel{formalT} $$
(cf. \eq\eqref{formalS} ). The explicit calculation of $\hat f^{(r)}_T$ is
difficult, but can be avoided by the following trick. Consider the
functions $f^{(r)}(\lambda)$ defined as
$$\eqalignno{
f^{(1)}(\lambda) &=
{h(\lambda)\over S^{(1)'}(0)} +
{g(\lambda + \alpha^{(1)} - \alpha^{(2)})\over S^{(2)'}(0)}
 \,,  \cr
f^{(2)}(\lambda) &=
{h(\lambda)\over S^{(2)'}(0)} +
{g(\lambda + \alpha^{(2)} - \alpha^{(1)})\over S^{(1)'}(0)}
\,, \eqalignnum \cr} $$
where $\lambda$ ranges over the entire real line. From Eqs. \eqref{fT1}
and \eqref{fT2}, it is evident that
$$f_T^{(r)}(\lambda) = f_+^{(r)}(\lambda) \,, \eqnum $$
where
$$ f_+(\lambda) \equiv \left\{ \matrix{ f(\lambda) & \lambda > 0 \cr
                                        0          & \lambda < 0 \cr} \right.
\,, \qquad\qquad
   f_-(\lambda) \equiv \left\{ \matrix{   0          & \lambda > 0 \cr
                                          f(\lambda) & \lambda < 0 \cr} \right.
\,, \qquad\qquad f = f_+ + f_- \,.
\eqnum $$
The Fourier transform of $f^{(r)}(\lambda)$ is readily computed, and can
be expressed in terms of the kernel $\hat K$,
$$\hat f(\omega) = \hat K(\omega)\ \left( \matrix{ 1/S^{(1)'}(0) \cr
                                                   1/S^{(2)'}(0) \cr} \right)
\,. \eqnum $$
{}From the factorization \eqref{factorization}, it follows that
$$\hat f_+ + \hat f_- = \left( 1 - G_-^{-1} G_+^{-1} \right)
\left( \matrix{ 1/S^{(1)'}(0) \cr
                1/S^{(2)'}(0) \cr} \right) \,. \eqnum $$
After multiplying both sides of this equation by $G_-$, we see that
the $P_+$ projection of $G_- \hat f_+$ is given by
$$P_+ \left( G_- \hat f_+ \right) = \alpha - G_+^{-1}
\left( \matrix{ 1/S^{(1)'}(0) \cr
                1/S^{(2)'}(0) \cr} \right) \,, \eqlabel{Q+} $$
where $\alpha$ is a constant. Requiring the right hand side to vanish
for $|\omega| \rightarrow \infty$ in $\Pi_+$ determines this constant to
be
$$\alpha =
\left( \matrix{ 1/S^{(1)'}(0) \cr
                1/S^{(2)'}(0) \cr} \right) \,. \eqnum $$
We conclude from \eqref{formalT} that $\hat T(\omega)$ is given by
$$\hat T (\omega) = \left( G_+(\omega) - 1 \right)
\left( \matrix{ 1/S^{(1)'}(0) \cr
                1/S^{(2)'}(0) \cr} \right) \,, \eqlabel{Tsolution} $$
where $S^{(r) '}(0)$ is given by \eqref{derivative}.

To summarize this section: we have made the expansion \eqref{expansion}
of $\epsilon_1^{(r)}(\lambda)$, and we have changed in Eqs. \eqref{funcS},
\eqref{funcT} from the variables
$\varepsilon^{(r)}(\lambda)$, $\eta^{(r)}(\lambda)$ to the variables
$S^{(r)}(\lambda)$, $T^{(r)}(\lambda)$, respectively. Using Wiener-Hopf
methods, we have determined in Eqs. \eqref{Ssolution}, \eqref{Tsolution} the
corresponding Fourier transforms $\hat S^{(r)}(\omega)$, $\hat T^{(r)}(\omega)$
in the limits $T \rightarrow 0$ and $H \rightarrow 0$. These expressions
involve $G_+(\omega)$ and $G_-(\omega)$, which appear in the factorization
\eqref{factorization}. In the next section, we shall use these results
to calculate the free energy.

\vfill\eject
\noindent
{\bf \chapnum . The free energy}
\vskip 0.2truein

Substituting the expansion \eqref{expansion} of $\epsilon_1^{(r)}$ into
the expression \eqref{free4} for the free energy (keeping in mind the
discussion immediately following \eqref{expansion}), we obtain
$${F\over N} = e_0
-{\pi^2 T^2\over 3}\sum_{r=1}^2 {s^{(r)}(\alpha^{(r)}) \over t^{(r)}}
- 2 \sum_{r=1}^2 \int_{\alpha^{(r)}}^\infty  d\lambda\ s^{(r)}(\lambda)
\left[  \varepsilon^{(r)}(\lambda) +  \eta^{(r)}(\lambda) \right]
\,, \eqlabel{freeint} $$
where the ground state energy per site $e_0$ is given by
$$e_0 = -2\pi \int_{-\infty}^\infty  d\lambda\  a_1 (\lambda)
s^{(1)}(\lambda) \,. \eqnum $$

We aim for a double expansion of $F/N$ to quadratic order in both $T$ and $H$.
To this end, we rewrite \eqref{freeint} in terms of $S^{(r)}(\lambda)$ and
$T^{(r)}(\lambda)$.
Consider first the $T^2$ term. For $H \rightarrow 0$, we can make
the approximation
$$ s^{(r)}(\alpha^{(r)}) = {1\over {\sqrt 3}} e^{-2\pi \alpha^{(r)}/3}
\,, \eqnum $$
and hence this term becomes
$$ -{\pi^2 T^2 \over 3{\sqrt 3}} \sum_{r=1}^2 {\kappa^{(r)} \over S^{(r)'}(0)}
\,. \eqnum $$
We next manipulate the integral in the last term as follows:
$$\eqalignno{
\int_{\alpha^{(r)}}^\infty & d\lambda\ s^{(r)}(\lambda)
\left[  \varepsilon^{(r)}(\lambda) +  \eta^{(r)}(\lambda) \right] \cr
& = \int_0^\infty  d\lambda\ s^{(r)}(\lambda + \alpha^{(r)})
\left[ {e^{-2\pi \alpha^{(r)}/3}\over \kappa^{(r)}} S^{(r)}(\lambda)
+  {\pi^2 T^2 \kappa^{(r)}
\over 6 e^{-2\pi \alpha^{(r)}/3}} T^{(r)}(\lambda) \right] \,.
\eqalignnum \cr}$$
Taking the $\alpha^{(r)} \rightarrow \infty$ limit of this expression,
we arrive at the following expression for $F/N$,
$$\eqalignno{
{F\over N} &= e_0
-{{\sqrt 3} H^2 \over 2\pi^2} \sum_{r=1}^2 \kappa^{(r)}
\int_0^\infty  d\lambda\ e^{-2\pi \lambda/3}  S^{(r)}(\lambda) \cr
& \qquad  - {\pi^2 T^2 \over 3{\sqrt 3}} \sum_{r=1}^2 \kappa^{(r)} \left[
\int_0^\infty  d\lambda\ e^{-2\pi \lambda/3} T^{(r)}(\lambda) +
{1\over S^{(r)'}(0)} \right] \,.
\eqalignnum  \cr}$$
Writing the functions $S^{(r)}(\lambda)$ and $T^{(r)}(\lambda)$
as the Fourier transforms of $\hat S(\omega)$ and $\hat T(\omega)$,
respectively, and then performing both the $\lambda$ and $\omega$
integrals, we obtain
$${F\over N} = e_0
-{{\sqrt 3} H^2 \over 2\pi^2} A  - {\pi^2 T^2 \over 3{\sqrt 3}} B \,, \eqnum $$
with
$$ \eqalignno{
A &= \sum_{r=1}^2  \kappa^{(r)} \hat S^{(r)}(2\pi i/3) =
\kappa^T\ \hat S(2\pi i/3) \,, \cr
B &= \sum_{r=1}^2  \kappa^{(r)} \left[ \hat T^{(r)}(2\pi i/3)
+ {1\over S^{(r)'}(0)} \right] =
\kappa^T\ \left[ \hat T(2\pi i/3)
+ \left( \matrix{ {1/ S^{(1)'}(0)} \cr
                  {1/ S^{(2)'}(0)} \cr}  \right) \right]  \,.
\eqalignnum  \cr}$$
where we again switch to a matrix notation.

Since both $A$ and $B$ involve $\kappa^T$, we begin by observing from
\eqref{explicitkappa} that
$$\kappa = \left( \matrix{ \kappa^{(1)} \cr
                           \kappa^{(2)} \cr} \right)
= G_-(-2\pi i/3)^{-1} G_-(0) \left( \matrix{ 1 \cr
                                             1 \cr} \right) \,.
\eqnum $$
Taking the transpose of this equation, and using the property
\eqref{Gproperty}, we obtain
$$\kappa^T = \left( \kappa^{(1)}\ \ \kappa^{(2)} \right)
= \left( 1\ \ 1 \right)\
G_+(0) G_+(2\pi i/3)^{-1}  \,. \eqnum $$

We now consider $A$. Evaluating $\hat S(2\pi i/3)$ using the expression
\eqref{Ssolution}, we obtain
$$ A = {{\sqrt 3}\over 2}\left( 1\ \ 1 \right)\
G_+(0) G_-(0) \left( \matrix{ 1 \cr
                              1 \cr} \right) \,.
\eqnum $$
Recalling the factorization equation \eqref{factorization} and the explicit
expression \eqref{Kkernel} for $\hat K(\omega)$, we conclude that
$$ A = {{\sqrt 3}\over 2}\left( 1\ \ 1 \right)\
\left( 1 - \hat K(0) \right)^{-1} \left( \matrix{ 1 \cr
                                                  1 \cr} \right)
= {\sqrt 3} \,. \eqnum $$

There remains to compute $B$. Evaluating $\hat T(2\pi i/3)$ using the
expression \eqref{Tsolution}, we obtain (after a crucial cancelation)
$$ B = \left( 1\ \ 1 \right)\ G_+(0)
\left( \matrix{ {1/ S^{(1)'}(0)} \cr
                {1/ S^{(2)'}(0)} \cr}  \right)   \,.
\eqnum $$
Recall Eq. \eqref{derivative},
$$\left( \matrix{ S^{(1)'}(0) \cr
                  S^{(2)'}(0) \cr}  \right) = {4 \pi^2 \over 3{\sqrt 3}}
G_-(0)
\left( \matrix{ 1 \cr
                1 \cr} \right) \,. \eqnum $$
Taking the transpose of this equation and again using the property
\eqref{Gproperty}, we see that
$$ B = {3{\sqrt 3}\over 4 \pi^2} \left( S^{(1)'}(0)\ \ S^{(2)'}(0) \right)\
\left( \matrix{ 1/S^{(1)'}(0) \cr
                1/S^{(2)'}(0) \cr}  \right) = {3{\sqrt 3}\over 2\pi^2} \,.
\eqnum $$

The expression for the free energy per site is therefore
$${F\over N} = e_0 - {3\over 2\pi^2} H^2 - {1\over 2}T^2 \,. \eqnum $$
As foreseen above, this result was obtained without using
explicit expressions for $G_+$ and $G_-$.
It follows that the magnetic susceptibility and specific heat, to lowest
order, are given by
$$\eqalignno{
\chi &= - {\partial^2\over \partial H^2} \left( {F\over N} \right)
\Big\vert_T = {3\over \pi^2} \,, \cr
C_H &= -T {\partial^2\over \partial T^2} \left( {F\over N} \right)
\Big\vert_H = T \,.  \eqalignnum \cr} $$

For a critical chain, the free energy per site is given
by${}^{\refref{cardy}, \refref{affleck1}}$
$${F\over N} = e_0 - {\pi c\over 6 v_s} T^2 + \cdots \,, \eqnum $$
where $c$ is the central charge and $v_s$ is the velocity of sound.
For our model, $v_s = 2\pi/3$ (see \eq\eqref{dispersion}), and hence $c=2$.

\vfill\eject
\noindent
{\bf \chapnum . Discussion}
\vskip 0.2truein

For an integrable model, the algebraic calculations leading to the energy
eigenvalues in terms of solutions of the BA equations are quite elegant
and precise. The same cannot be said for the corresponding thermodynamic
calculations, at least in their present formulation. This is both surprising
and disappointing. For instance, one expects that there should be an
analogue of the Sugawara construction for integrable models, by which
one could determine the low-temperature specific heat (central charge)
by purely algebraic means. The fact that an explicit Wiener-Hopf factorization
is not needed to compute such properties also suggests that an alternative
approach should be possible. Indeed, very recently, progress has been
made${}^{\refref{jimbo}}$ towards an algebraic formulation of the
thermodynamics
of integrable models.

Despite its short-comings, the approach which we have followed here to
investigate the low-temperature thermodynamics of the closed $su(3)$-invariant
chain in the fundamental representation is nevertheless practical. It
reproduces the known result for the central charge, and evidently, it can be
implemented for $su(n)$.

We recall${}^{\refref{affleck1}}$ that, in the continuum limit, integrable
spin-$s$ $su(2)$-invariant chains${}^{\refref{xxx/spin/s}}$ are described by
level $k=2s$ $su(2)$ WZW models. Moreover, Affleck has
found${}^{\refref{affleck2}}$ a simple
relation between the magnetic susceptibility $\chi$ and the level, namely
$\chi v_s = k/2\pi$. Presumably, there is a generalization of this relation to
the $su(n)$ case, for which there are $n-1 ( =$ rank of $su(n) )$ magnetic
susceptibilities $\chi_i$, $i= 1, \cdots, n-1$. We have determined for
$su(3)$  a particular linear combination of magnetic susceptibilities, which
is dictated by the special imbedding $so(3) \subset su(3)$ which characterizes
the $A^{(2)}_2$ chain. Our result may be relevant to the
$su(n)$-generalization of Affleck's relation.

Having treated the noncritical regime of the $A^{(2)}_2$ chain, the task
now is to investigate the critical regime. As noted in the Introduction,
this may be more feasible for the open chain, which has $U_q [su(2)]$
symmetry. We hope to report on this problem in the future.

\bigskip

We thank A. Jacob for bringing Ref. \refref{gohberg/krein} to our attention,
and E. Melzer for useful discussions.
Part of this work was performed at the Aspen Center for Physics.
This work was supported in part by the National Science
Foundation under Grants No. PHY-90 07517 and PHY-92 09978.

\vfill\eject

\noindent
{\bf References}

\vskip 0.2truein

\reflabel{qism}
R.J. Baxter, {\it Exactly Solved Models in Statistical Mechanics}
(Academic Press, 1982);
L.D. Faddeev and L.A. Takhtajan, Russ. Math. Surv. {\it 34} (1979) 11;
P.P. Kulish and E.K. Sklyanin, J. Sov. Math. {\it 19} (1982) 1596;
{\it Lecture Notes in Physics} {\it 151} (Springer, 1982) 61.

\reflabel{yangyang1}
C.N. Yang and C.P. Yang, J. Math. Phys. {\it 10} (1969) 1115. For an
excellent review, see N.M. Bogoliubov, A.G. Izergin and V.E. Korepin, in
{\it Lecture Notes in Physics} {\it 242} (Springer, 1985) 220.

\reflabel{takahashi1}
M. Takahashi, Prog. Theor. Phys. {\it 46} (1971) 401.

\reflabel{gaudin}
M. Gaudin, Phys. Rev. Lett. {\it 26} (1971) 1301;
{\it La fonction d'onde de Bethe} (Masson, 1983).

\reflabel{johnson/mccoy}
J.D. Johnson and B.M. McCoy, Phys. Rev. {\it A6} (1972) 1613.

\reflabel{takahashi/suzuki}
M. Takahashi and M. Suzuki, Prog. Theor. Phys. {\it 48} (1972) 2187.

\reflabel{takahashi2}
M. Takahashi, Prog. Theor. Phys. {\it 50} (1973) 1519;
{\it 51} (1974) 1348.

\reflabel{filyov}
V.M. Filyov, A.M. Tsvelik and P.B. Wiegmann, Phys. Lett. {\it 81A}
(1981) 175.

\reflabel{tsvelick/wiegmann}
A.M. Tsvelick and P.B. Wiegmann, Adv. in Phys. {\it 32} (1983) 453.

\reflabel{andrei}
N. Andrei, K. Furuya and J.H. Lowenstein, Rev. Mod. Phys. {\it 55}
(1983) 331.

\reflabel{faddeev}
L.D. Faddeev and L.A. Takhtajan, J. Sov. Math. {\it 24} (1984) 241.

\reflabel{xxx/spin/s}
H.M. Babujian, Nucl. Phys. {\it B215} (1983) 317;
L.A. Takhtajan, Phys. Lett. {\it 87A} (1982) 479;

\reflabel{xxz/spin/s}
K. Sogo, Phys. Lett. {\it 104A} (1984) 51;
H.M. Babujian and A.M. Tsvelick, Nucl. Phys. {\it B265} (1986) 24;
A.N. Kirillov and N.Yu. Reshetikhin, J. Phys. {\it A20} (1987) 1565;
J. Sov. Math. {\it 35} (1986) 2627.

\reflabel{kingston}
L. Mezincescu and R.I. Nepomechie, in {\it Quantum Groups, Integrable Models
and Statistical Systems}, ed. by J. Le Tourneux and L. Vinet (World
Scientific), in press.

\reflabel{izergin/korepin}
A.G. Izergin and V.E. Korepin, Commun. Math. Phys. {\it 79} (1981) 303.

\reflabel{reshetikhin1}
V.I. Vichirko and N.Yu. Reshetikhin, Theor. Math. Phys. {\it 56}
(1983) 805; N.Yu. Reshetikhin, Lett. Math. Phys. {\it 7} (1983) 205.
N.Yu. Reshetikhin, Sov. Phys. JETP {\it 57} (1983) 691.

\reflabel{tarasov}
V.O. Tarasov, Theor. Math. Phys. {\it 76} (1988) 793.

\reflabel{nienhuis1}
B. Nienhuis, Int. J. Mod. Phys. {\it B4} (1990) 929.

\reflabel{reshetikhin2}
N. Yu. Reshetikhin, J. Phys. {\it A24} (1991) 2387.

\reflabel{nienhuis2}
B. Nienhuis, Phys. Rev. Lett. {\it 49} (1982) 1062.

\reflabel{batchelor1}
M.T. Batchelor, B. Nienhuis and S.O. Warnaar,
Phys. Rev. Lett. {\it 62} (1989) 2425.

\reflabel{miami}
L. Mezincescu and R.I. Nepomechie, in {\it Quantum Field Theory,
Statistical Mechanics, Topology and Quantum Groups}, ed. by T.L. Curtright,
L. Mezincescu and R.I. Nepomechie (World Scientific, 1992) 200.

\reflabel{devega1}
H.J. de Vega and E. Lopes, J. Phys. {\it A23} (1990) L905 and Erratum;
Nucl. Phys. {\it B362} (1991) 261.

\reflabel{batchelor2}
S.O. Warnaar, M.T. Batchelor and B. Nienhuis, J. Phys. {\it A25} (1992)
3077.

\reflabel{ijmp}
L. Mezincescu and R.I. Nepomechie, Int. J. Mod. Phys. {\it A6} (1991) 5231;
Addendum, {\it A7} (1992) 5657;
Mod. Phys. Lett. {\it A6} (1991) 2497.

\reflabel{analytical}
L. Mezincescu and R.I. Nepomechie, Nucl. Phys. {\it B372} (1992) 597.

\reflabel{pasquier/saleur}
V. Pasquier and H. Saleur, Nucl. Phys. {\it B330} (1990) 523.

\reflabel{sutherland}
B. Sutherland, Phys. Rev. {\it B12} (1975) 3795.

\reflabel{kulish/reshetikhin}
P.P. Kulish and N.Yu. Reshetikhin, Sov. Phys. JETP {\it 53} (1981) 108.

\reflabel{yangyang2}
C.N. Yang and C.P. Yang, Phys. Rev. {\it 150} (1966) 327.

\reflabel{gohberg/krein}
I.C. Gohberg and M.G. Krein, in {\it American Mathematical Society
Translations, Series 2, Vol 14} (American Mathematical Society, 1960) 217.
See also M.G. Krein, in {\it American Mathematical Society
Translations, Series 2, Vol 22} (American Mathematical Society, 1962) 163.

\reflabel{cardy}
H.W.J. Bl\"ote, J.L. Cardy and M.P. Nightingale, Phys. Rev. Lett. {\bf 56}
(1986) 742.

\reflabel{affleck1}
I. Affleck, Phys. Rev. Lett. {\bf 56} (1986) 746.

\reflabel{finite-size}
S.V. Pokrovskii and A.M. Tsvelik, Sov. Phys. JETP {\it 66} (1987) 1275;
H.J. de Vega, J. Phys. {\it A20} (1987) 6023.

\reflabel{jimbo}
B. Davies, O. Foda, M. Jimbo, T. Miwa, and A. Nakayashiki, RIMS preprint;
M. Jimbo, K. Miki, T. Miwa, and A. Nakayashiki, RIMS preprint.

\reflabel{affleck2}
I. Affleck, Phys. Rev. Lett. {\it 56} (1986) 2763.

\end